%% file: paper.tex
\documentclass[sigconf,screen,authorversion]{acmart}
\usepackage{cleveref}
\usepackage{xspace}
\usepackage{blindtext}
\usepackage{framed}
\usepackage{fancyvrb}
\usepackage{caption}
\usepackage{subcaption}
\usepackage{xcolor}
\usepackage{listings}
\usepackage{enumitem}

\lstdefinelanguage{IR} {
  basicstyle=\ttfamily,
  mathescape=true,
  emphstyle={\textbf},
  emph={let,in},
}

\newcommand{\term}[1]{\texttt{#1}}

\crefname{section}{\S}{\S\S}
\Crefname{section}{\S}{\S\S}

\copyrightyear{2022} 
\acmYear{2022} 
\setcopyright{acmlicensed}\acmConference[ICSE-NIER'22]{New Ideas and Emerging Results }{May 21--29, 2022}{Pittsburgh, PA, USA}
\acmBooktitle{New Ideas and Emerging Results (ICSE-NIER'22), May 21--29, 2022, Pittsburgh, PA, USA}
\acmPrice{15.00}
\acmDOI{10.1145/3510455.3512787}
\acmISBN{978-1-4503-9224-2/22/05}

\begin{document}

\title{Grammars for Free: Toward~Grammar~Inference~for~Ad~Hoc~Parsers}

\author{Michael Schröder}
\orcid{0000-0003-1496-0531}
\affiliation{%
  \institution{TU Wien}
  \city{Vienna}
  \country{Austria}
}
\email{michael.schroeder@tuwien.ac.at}

\author{Jürgen Cito}
\affiliation{%
  \institution{TU Wien and Meta Platforms, Inc.}
  \city{Vienna}
  \country{Austria}
}
\email{juergen.cito@tuwien.ac.at}

\authorsaddresses{}

\begin{abstract}
  Ad hoc parsers are everywhere: they appear any time a string is split, looped over, interpreted, transformed, or otherwise processed.
  Every ad hoc parser gives rise to a language: the possibly infinite set of input strings that the program accepts without going wrong.
  Any language can be described by a formal grammar: a finite set of rules that can generate all strings of that language.
  But programmers do not write grammars for ad hoc parsers---even though they would be eminently useful.
  Grammars can serve as documentation, aid program comprehension, generate test inputs, and allow reasoning about language-theoretic security.
  We propose an automatic grammar inference system for ad hoc parsers that would enable all of these use cases, in addition to opening up new possibilities in mining software repositories and bi-directional parser synthesis.
\end{abstract}

\maketitle

\section{Introduction}
\label{sec:intro}

\emph{Parsing} is one of the fundamental activities in software engineering.
Following \citet{grunejacobs2008}, we take parsing to mean ``the process of structuring a linear representation in accordance with a given grammar,'' an activity so common that pretty much every program performs some kind of parsing at one point or another.
Academically, parsing has been studied since the very early days of computer science \cite{irons1961algol} and \emph{formal language theory}, which has its origin in linguistics \cite{chomsky1957syntactic}, provides the foundation for an impressive amount of both theoretical results \cite{hopcroft1979intro} and practical applications \cite{grunejacobs2008}.
As part of every-day programming, \emph{regular expressions} \cite{thompson1968regex} are probably the biggest and most widely known success story of applied formal language theory.
But apart from regexes, only a small minority of programs, mainly compilers and some protocol implementations, make explicit reference to the formal-theoretic underpinnings of parsing, documenting grammars of their input languages and making use of formalized parsing techniques such as parser generators \cite{johnson1990yacc,parr1995antlr} or parser combinator frameworks \cite{leijen2001parsec}.
The vast majority of parsing code in software today is \emph{ad hoc}.

\begin{figure}
\begin{framed}
\begin{BVerbatim}
xs = map(int, s.split(','))
\end{BVerbatim}
\vspace{1em}
\hrule
\begin{align*}
  s\ &\to\ int \mid int\ \term{,}\ s\\
  int\ &\to\ space^*\ sign^?\ digit\ (\term{\_}^?\ digit)^*\ space^*\\
  digit\ &\to\ \term{0} \mid \term{1} \mid \term{2} \mid \term{3} \mid \term{4} \mid \term{5} \mid \term{6} \mid \term{7} \mid \term{8} \mid \term{9}\\
  sign\ &\to\ \term{+} \mid \term{-}\\
  space\ &\to\ \term{\textvisiblespace} \mid \term{\textbackslash t} \mid \term{\textbackslash n} \mid \term{\textbackslash v} \mid \term{\textbackslash f} \mid \term{\textbackslash r}
\end{align*}
\vspace{0.1em}
\hrule
\vspace{1em}
\def\svgwidth{\textwidth}
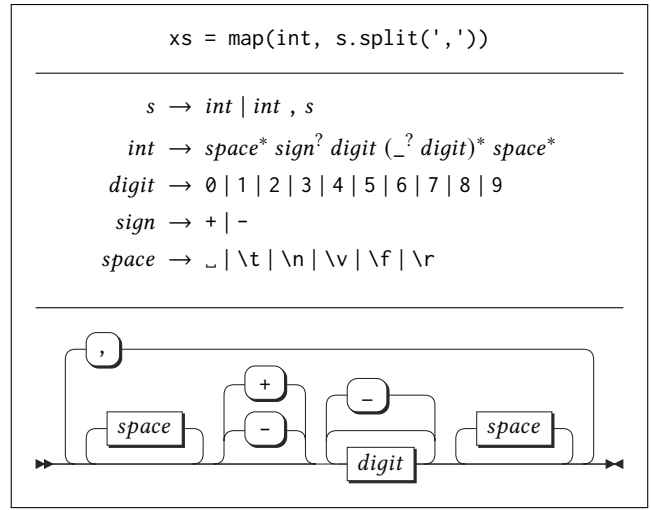
\end{framed}
  \vspace{-1em}
  \caption{An ad hoc parser and its grammar.\protect\footnotemark}
  \label{fig:parser}
  \vspace{-1.5em}
\end{figure}

The Python expression in \Cref{fig:parser} is a typical example of an \emph{ad hoc parser}.
It transforms a string \texttt{s} into a list of integers \texttt{xs}.
First, the \texttt{split} function breaks \texttt{s} into its comma-separated substrings, then the \texttt{map} function applies the \texttt{int} constructor to all substrings, turning each into a proper integer value.
This parser does not use any particular parsing techniques or frameworks, just ordinary functions manipulating strings and transforming values.
A programmer writing this expression would most likely not think about the fact that they are writing a parser.
Splitting a comma-separated list of values, just like extracting a command-line argument, reading a timestamp, or any other minor programming task involving strings, barely registers as parsing.
Commonly, this kind of parsing code is deeply entangled with application logic---a phenomenon known as \emph{shotgun parsing} \cite{momot2016seven}.

\footnotetext{The notation used here denotes \term{terminals} with typewriter font and uses the common operators $*$, $+$, and $?$ for zero-or-more, one-or-more, and optional occurrences, respectively. The vertical bar $\mid$ separates alternative productions, without precedence. Parentheses are used for grouping in the usual way.}

\Cref{fig:parser} also includes a complete grammar for this parser (assuming the semantics of Python 3.9).
It is not a particularly complex grammar, but it is perhaps still surprising.
Even an experienced Python programmer might be unaware, for example, that the \texttt{int} constructor, in addition to allowing an optional leading \texttt{+} or \texttt{-} sign, also permits leading zeroes, strips surrounding whitespace, and ignores single \texttt{\_} characters that are used for grouping digits.
Looking at the grammar, we can see that the strings \texttt{"12,304"} and \texttt{"+01\_2,3\_0\_4\textvisiblespace"} will both be accepted by the parser, while the empty input \texttt{""} will crash the program.
That the parser's language excludes the empty string is obvious from the grammar, but might be difficult to work out from looking at the code alone.%
\footnote{The \texttt{split} function, when applied to an empty string, returns a singleton list also containing an empty string (rather than an empty list, as one might assume). The \texttt{int} constructor, applied to this empty string via \texttt{map}, will then throw a runtime exception.}

A grammar certainly reveals a great deal about a rather deceptively simple looking expression, yet no programmer would actually write it down.
Grammars share the same fate as most other forms of specification: they are hard to write, can be hard to read, and seem hardly worth the trouble---especially for ad hoc parsers.
\emph{If we are not building whole houses, why should we draw blueprints?} \cite{lamport2015builds}

But there is a form of specification, one wildly more successful than grammars, that we can draw inspiration from: \emph{types}.
Formal grammars are similar to types, in that a parser without a grammar is very much like a function without a type signature.
Types have one significant advantage over grammars, however: most type systems offer a form of \emph{type inference}, allowing programmers to omit type annotations because they can be automatically recovered from the surrounding context.\footnote{For a good introduction to type inference and its history, see \cite[\S~4]{macqueen2020hopl}.}
If we could infer grammars like we can infer types, we could reap all the rewards of having a complete specification of our program's input language, without burdening the programmer with the full weight of formal language theory.

In this work, we sketch a possible path towards inferring grammars for ad hoc parsers by combining methods found in refinement types and string constraint solving. 
We describe future possibilities where grammar inference enables, among other things, better program comprehension by explicitly documenting a program's input space, and bi-directional parser synthesis that helps developers refine and secure their input validation.

\section{The Need for Grammars}
\label{sec:need}

Before we delineate how to statically infer grammars, we want to briefly motivate why every (ad hoc) parser would greatly benefit from having a known grammar. 

\paragraph{\textbf{Documentation}}
A formal grammar is the ideal documentation for a parser, because it provides a high-level perspective that focuses on the \emph{data} as opposed to the code.
It allows the programmer to grasp the input language as is, without being distracted by the mechanics of the implementation.
There exist numerous notations for grammars, each suitable for different languages and in different contexts:
regular expressions~\cite{thompson1968regex}, Chomsky normal form~\cite{chomsky1959}, Augmented Backus-Naur Form (ABNF)~\cite{rfc5234}, parsing expression grammars (PEGs)~\cite{ford2004peg}, etc.
Graphic representations, like finite state machines~\cite{hopcroft1979intro} or railroad diagrams~\cite{braz1990visual} (see~\Cref{fig:parser}), can be particularly helpful in understanding abstract data
and align with developers' appreciation of sketches and diagrams \citep{baltes2014sketches}.

\paragraph{\textbf{Program Comprehension}}
It is known that providing alternative representations for a programming task can increase program comprehension \cite{gilmore1984comprehension,fitter1979diagrams}.
The example in \Cref{fig:parser} demonstrates how a grammar can elucidate the corresponding ad hoc parsing code, revealing otherwise hidden features and potentially bugs or security issues.
Because a grammar is also a \emph{generating device}, it is possible to construct any sentence of its language in a finite number of steps---manually or in an automated fashion.
Generating concrete examples of possible inputs further helps in understanding parsing code, and can be invaluable during testing and debugging.

\paragraph{\textbf{Fuzzing}}
We can test programs by bombarding them with (systematically generated) random inputs and seeing if anything breaks.
This is known as fuzz testing, or fuzzing \cite{fuzzingbook2021,manes2019art}.
Generating good fuzz inputs is not easy, because in order to penetrate into deep program states, one generally needs valid or near-valid inputs, meaning inputs that pass at least the various syntactic checks and transformations---i.e. ad hoc parsers---scattered throughout a typical program.
One promising approach is \emph{grammar-based fuzzing} \cite{holler2012fuzzing,aschermann2019nautilus}, where valid inputs are specified with the help of language grammars.

\paragraph{\textbf{Language-Theoretic Reasoning}}
As formal descriptions of input languages, grammars allow us to reason about various language-theoretic properties, such as computability bounds. The \emph{language-theoretic security} (LANGSEC) community\footnote{\url{https://langsec.org}} regards such reasoning as vital in assuring the correctness and safety of input handling routines.
For example, if an input language is recursively enumerable, we can never guarantee that its parser behaves safely (i.e. halts) on inputs that are not in the language, because the parser must be equivalent to a Turing machine.
Thus, input languages should be minimally powerful, and their parsers should match them in computational power \cite{sassaman2013langsec}.
Ad hoc parsers open themselves up to attack, because it is not clear what languages they implement, or if they implement them correctly, and variations among implementations are easily overlooked \cite{schneeweisz2020differentials}.
Grammars can help assure us that our input languages have favorable properties and that their parsers are implemented correctly. 

\paragraph{\textbf{Automatic Parser Generation}}
A \emph{parser generator} is a tool that synthesizes a parser from a given grammar.
Examples include Yacc~\citep{johnson1990yacc}, ANTLR~\citep{parr1995antlr}, and OMeta~\citep{warth2007ometa}.
These tools are common in certain areas, such as compilers, and are usually invoked during program build time, generating parsing code that is linked with the rest of the program.
The great advantage of starting with a grammar and letting the parser implementation be generated automatically is a high assurance of correctness, as well as easier maintainability.

\section{Toward Grammar Inference}
\label{sec:inference}

We hope to realize automatic grammar inference based on the following intuition: 
Any parser is essentially a \emph{machine} in the formal sense---it is a recognizer for its input language.

\subsection{Background: Languages \& Machines}
Formally, a \emph{language} $L$ is a possibly infinite set of sentences over a finite alphabet $\Sigma$.
We can define languages very abstractly, as in $L = \{a^nb^n \mid n>0 \}$, a language over the alphabet $\Sigma=\{a,b\}$ that consists of all sentences with at least one $a$ followed by the same number of $b$s.
Usually, however, we define languages via generative devices called \emph{grammars} or recognizing devices called \emph{machines}.

A grammar $G=(V, \Sigma, P, S)$ is a finite description of a language and consists of a set of \emph{variables} (or \emph{nonterminals}) $V$; a \emph{terminal} alphabet $\Sigma$; a set of \emph{productions} $P$, which are rules of the form $\alpha\to\beta$ where $\alpha$ and $\beta$ are from $V$ and/or $\Sigma$; and a \emph{start symbol} $S\in V$.
By starting with $S$ and applying a finite number of productions from $P$, we can generate sentences over $\Sigma$.
The language $L(G)$ is the set of all sentences that can be generated by $G$.
By putting various constraints on the \emph{form} of a grammar, such as whether the left-hand side of a production can only include variables, or the right-hand side has to include at least one terminal symbol, and so on, we can limit the grammar's expressiveness, constraining the \emph{family} of languages a grammar of this form can produce.
The famous {Chomsky hierarchy}~\cite{chomsky1959} partitions languages/grammars into four increasingly expressive levels: \emph{regular}, \emph{context-free}, \emph{context-sensitive}, and \emph{recursively enumerable}.
Numerous additional language families and types of grammars have been discovered, within and beyond the classic hierarchy: \emph{attribute grammars} \cite{knuth1968semantics}, \emph{boolean grammars}~\cite{okhotin2004boolean,okhotin2013conjunctive}, the \emph{mildly context-sensitive} and \emph{sub-regular} languages \cite{jager2012formal}, \emph{parsing expression grammars} (PEGs)~\cite{ford2004peg}, to name just a few.

A machine $M$, unlike a grammar, does not produce sentences but consumes them.
Taking some sentence as input and moving through a finite number of internal states, it arrives at some halting configuration if and only if the sentence is part of the language $L(M)$.
If the sentence is not part of the language, the machine either runs forever or gets stuck in a non-accepting state.
Just like with grammars, the way that a machine is constructed determines its expressiveness.
There is a natural correspondence between languages, grammars, and machines: 
regular languages correspond to \emph{finite state machines}, which simply move from one internal state to another based on the next input character; 
context-free languages correspond to finite state machines equipped with a pushdown stack, also known as \emph{pushdown automata};
context-sensitive languages correspond to \emph{linearly bounded automata}, in essence Turing machines with a finite tape;
and finally the recursively enumerable languages correspond to the well-known unbounded \emph{Turing machines}.
As with grammars, there are numerous additional and alternative constructions between and beyond these classic ones.

\begin{figure}
  \centering
  \includegraphics[width=0.95\columnwidth]{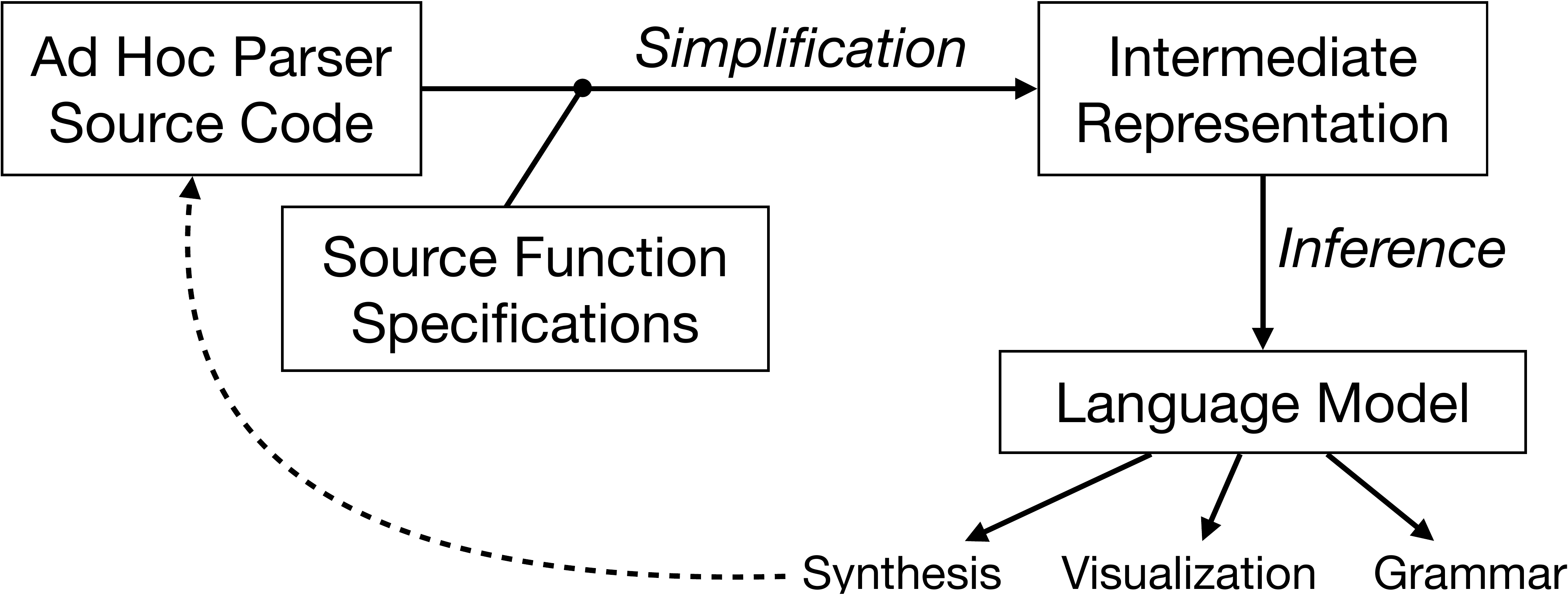}
  \caption{Sketch of our grammar inference system.}
  \label{fig:sketch}
  \vspace{-1.5em}
\end{figure}

\subsection{Intuition: Parsers are Embedded Machines}
A parser, like the Python snippet in \Cref{fig:parser}, which expects a string from some language $L_1$ as input, can be seen as a machine $M_1$ recognizing that language, so that $L_1=L(M_1)$.
This machine is however embedded within the more powerful machine $M_0$, the general-purpose programming language that the parser itself is written in.
Any real world parser will do more than just recognize a language:
it will allocate and transform data types, throw exceptions or handle parse errors, or perform side effects unrelated to the parsing process itself.
Nevertheless, the control flow at the core of a parser will, in our experience, closely match that of the (hypothetical) machine $M_1$.
While it is entirely possible that a parser written in a Turing-complete programming language exhibits exactly those traits that make it equivalent to a Turing machine, even though it might be parsing a ``lesser'' language, we think this to be very unlikely.
In almost all practical situations, ad~hoc parsing code will not significantly exceed the ``power-level'' of the language it is parsing.
For example, unless it has been especially constructed to be confounding, the loops present in a parser will invariably be bounded by at most some linear factor of the length of its input, which corresponds to the expressiveness of a context-sensitive language.
Thus, we think it is feasible to transform ad hoc parsers into equivalent machines whose languages can be statically inferred.

\subsection{Vision: Automatic Grammar Inference}
\Cref{fig:sketch} shows a sketch of the end-to-end grammar inference system that we envision.
In the first step, ad hoc parsing code is transformed from a Turing-complete source language (e.g. Python) into an intermediate representation (IR) that is essentially a domain-specific language for parsing.
This transformation can be seen as a simplification: it removes syntactic sugar, makes control flow explicit, and throws away all parts of the source code that are not related to parsing.
During this step, known string processing functions are translated into one or more equivalent functions of the IR that precisely model the semantics of the source.
To illustrate, let us consider a slightly extended version of the example from \Cref{fig:parser}:
\begin{lstlisting}[basicstyle=\ttfamily,numbers=left,numberstyle=\color{black!33},xleftmargin=2em]
def vector_length(s):
  [x,y,z] = map(int, s.split(','))
  return math.sqrt(x**2 + y**2 + z**3)
\end{lstlisting}
The simplification results in roughly the following IR:
\begin{lstlisting}[language=IR,numbers=left,numberstyle=\color{black!33},xleftmargin=2em]
let parse = $\lambda$(s : String {$\star$}).
  let $\nu_1$ = split$_{\texttt{py}}$ "," s in
  let xs = map int$_{\texttt{py}}$ $\nu_1$ in
  let $\nu_2$ = length xs in
  let $\nu_3$ = equals 3 $\nu_2$ in
  assert $\nu_3$
\end{lstlisting}
Note that this function does not actually return anything.
The goal here is not to run it and obtain a result, but to fill the hole ($\star$) in its input type by inferring the appropriate string constraints.
To this end, the functions $\texttt{split}_\texttt{py}$ and $\texttt{int}_\texttt{py}$ precisely model their Python counterparts and refine their input and output types by imposing the constraints resulting from their modeled string processing behavior.
Note also how a remnant of the pattern match \texttt{[x,y,z]} from the source is present in form of an (indirect) constraint on the length of the string (lines 4--6 in the IR).

Inferring the type of \texttt{parse} and solving its string constraints results in a model of the original ad hoc parser's input language.
To make the resulting grammar traceable to the originating code, the model also contains rich source location information, which has to be threaded through both the simplification and inference steps.
In a final step, the language model can then be used to generate the desired textual, visual, and interactive grammar representations.

\section{Related Work}
\label{sec:related}

\paragraph{Grammatical Inference}
A related but different problem to our goal of finding a grammar given a parser is to find a grammar given a set of sentences.
This is known as \emph{grammatical inference} or \emph{grammar induction}.
Early results in computational linguistics quickly established fundamental limits of what could be achieved:
it was shown that not even regular languages can be identified given only positive examples \cite{gold1967limit}.
Nevertheless, with applications ranging from speech recognition to computational biology, grammatical inference is an active and vibrant field \cite{delahiguera2005inference,delahiguera2010book}.

\paragraph{Fuzzing}
A big problem in grammar-based fuzzing (cf. \cref{sec:need}) is obtaining accurate grammars or language models.
Black-box approaches try to infer a language model by poking the program with seed inputs and monitoring its runtime behavior \cite{bastani2017glade,godefroid2017fuzz}.
This has some theoretical limits \cite{angluin1987queries,angluin1995queries} and the amount of necessary poking (i.e. membership queries) grows exponentially with the size of the grammar.
White-box approaches make use of the program code and can thus use more sophisticated analysis techniques, e.g. taint tracking to monitor data flow between variables \cite{hoschele2016autogram} or tracking dynamic control flow and observing character accesses of input strings \cite{gopinath2020mimid}.
These approaches rely on dynamic execution, but can produce fairly accurate and human-readable grammars, at least in test settings.
They can not, however, provide any guarantees of correctness, and thus it is not possible to determine how accurate the resulting grammars really are.
In our vision, grammars are \emph{statically} inferred from source code and are always \emph{sound}.
By not relying on dynamic execution of whole programs, grammars can be extracted from individual functions or even partial programs, and it is not necessary to generate seed inputs to bootstrap inference.

\paragraph{String Constraint Solving}
String constraints are relations defined over string variables and arise out of program statements that manipulate strings, e.g. concatenation or substring replacement.
Reasoning about strings requires solving combinatorial problems involving such constraints, which is difficult to do both efficiently and completely, and a large number of approaches have been developed \cite{amadini2021survey,stanford2021constraints}.
Our problem of grammar inference is in some ways the inverse:
instead of wanting to model all possible strings a function can return or express, we want to model all possible strings a function can accept (without throwing an error or getting stuck).

\section{New Possibilities}
\label{sec:possibilities}

The end-to-end grammar inference system we envision (\cref{sec:inference}) will not only let us enjoy all the benefits that formal grammars provide in general (\cref{sec:need}), it also enables some exciting new possibilities.

\paragraph{\textbf{Interactive Documentation}}
A grammar that is automatically inferred will always be up-to-date---a significant advantage over manually written documentation, which tends to quickly drift from the object it documents \citep{lethbridge2003documentation}.
Furthermore, an inferred grammar could be closely linked directly to the underlying source code, making productions traceable to their origins.
One can imagine an interactive environment where hovering over parts of a grammar highlights the corresponding pieces of code---or even allows changing them by manipulating the high-level representation.

\paragraph{\textbf{Bi-directional Parser Synthesis}}
Combining grammar inference with parser generation enables a framework of bi-directional parser synthesis.
In the most basic case, starting from an existing complete parser implementation, the synthesizer can be used to generate different implementations according to certain criteria, e.g. performance or code style, by transformation via the inferred grammar---a specialized type of semantic program transformation \citep{cousot2002transformation}.
If the initial parser is incomplete, a bi-directional parser synthesizer can be used for \emph{program sketching} \citep{solarlezama2008sketching,lubin2020sketching,polikarpova2016synthesis}, wherein an initial implementation (a ``sketch'') is the basis of an initial grammar which can be manipulated by the user on a high level---perhaps graphically---to then in turn synthesize a completed or refined implementation.
If the sketch-synth loop can be sufficiently shortened, it can be the basis for a direct manipulation bi-directional programming system \citep{mayer2018bidirectional,chugh2016direct}, although based on transformations of the (specification of) inputs to the program rather than its outputs.

\paragraph{\textbf{Mining \& Learning}}
An inferred grammar abstracts over the underlying concrete parser implementation and can be viewed as an equivalence class, allowing us to group together different parser implementations with similar semantics.\footnote{While there are a number of theoretical bounds regarding the decidability of properties about grammars, it is in fact possible to efficiently decide equivalence for many types of grammars encountered in practice \citep{madhavan2015grammar}.}
This opens up new possibilities in mining software repositories, such as grammar-enhanced semantic code search \citep{mishne2012search,garciacontreras2016browsing,premtoon2020search} or detecting code clones \citep{juergens2009clones,yu2019clones} of ad hoc parsers.
By automatically inferring grammars for each code change, it also becomes possible to learn how (implicit) input specifications evolve over time, enabling a type of grammar-aware semantic change tracking \citep{raghavan2004dex,hanam2019change}.
Augmenting the code review process with current as well as historical grammar information would allow developers to be alerted when a code change introduces a perhaps unexpected change in input grammar.

\section{Future Plans}
\label{sec:future}

We want to build the grammar inference system described in this paper and apply it in real-world situations.
We plan to realize our vision in a series of upcoming works:
\begin{itemize}[leftmargin=0.3cm]
  \item We are currently conducting a mining study of ad hoc parsers in the wild, collecting common coding patterns in order to determine the possible scope of our system.
  \item We are currently investigating the use of refinement types \citep{jhala2020refinement} in combination with string constraint solving to realize inference of a language model from a simplified parsing IR. While we have seen initial success with smaller examples, we need to expand to more kinds of parsers to understand our scope and limitations.
  \item To ensure the validity of our approach, both simplification and inference need to be proven sound. We plan on supplying machine-checked proofs for both of these steps.
  \item To ensure the effectiveness of our approach, we plan on evaluating the system on a corpus of curated ad hoc parser samples from the real world. We have begun collection of a suitable dataset.
  \item We plan on conducting a large-scale mining study of inferred grammars, to demonstrate the usefulness of our system to applications of code mining and learning.
  \item We plan on conducting a number of user studies on grammar comprehension in order to determine the benefits and drawbacks of different textual and visual grammar representations.
\end{itemize}
We are excited about the prospects of automated grammar inference and invite the community to collaborate with us to realize our vision of ``grammars for free''.

\bibliographystyle{ACM-Reference-Format}
\bibliography{references}

\end{document}

%% file: rr_tight.pdf_tex
\begingroup%
  \makeatletter%
  \providecommand\color[2][]{%
    \errmessage{(Inkscape) Color is used for the text in Inkscape, but the package 'color.sty' is not loaded}%
    \renewcommand\color[2][]{}%
  }%
  \providecommand\transparent[1]{%
    \errmessage{(Inkscape) Transparency is used (non-zero) for the text in Inkscape, but the package 'transparent.sty' is not loaded}%
    \renewcommand\transparent[1]{}%
  }%
  \providecommand\rotatebox[2]{#2}%
  \newcommand*\fsize{\dimexpr\f@size pt\relax}%
  \newcommand*\lineheight[1]{\fontsize{\fsize}{#1\fsize}\selectfont}%
  \ifx\svgwidth\undefined%
    \setlength{\unitlength}{422.25bp}%
    \ifx\svgscale\undefined%
      \relax%
    \else%
      \setlength{\unitlength}{\unitlength * \real{\svgscale}}%
    \fi%
  \else%
    \setlength{\unitlength}{\svgwidth}%
  \fi%
  \global\let\svgwidth\undefined%
  \global\let\svgscale\undefined%
  \makeatother%
  \begin{picture}(1,0.25754885)%
    \lineheight{1}%
    \setlength\tabcolsep{0pt}%
    \put(0,0){\includegraphics[width=\unitlength,page=1]{rr_tight.pdf}}%
    \put(0.13943162,0.08614565){\color[rgb]{0.05098039,0.05098039,0.05098039}\makebox(0,0)[lt]{\lineheight{1.25}\smash{$space$}}}%
    \put(0,0){\includegraphics[width=\unitlength,page=2]{rr_tight.pdf}}%
    \put(0.37388988,0.1562984){\color[rgb]{0.03921569,0.03921569,0.03921569}\makebox(0,0)[lt]{\lineheight{1.25}\smash{\,\texttt{+}}}}%
    \put(0,0){\includegraphics[width=\unitlength,page=3]{rr_tight.pdf}}%
    \put(0.37388988,0.07814565){\color[rgb]{0.03921569,0.03921569,0.03921569}\makebox(0,0)[lt]{\lineheight{1.25}\smash{\,\texttt{-}}}}%
    \put(0,0){\includegraphics[width=\unitlength,page=4]{rr_tight.pdf}}%
    \put(0.54795737,0.02275488){\color[rgb]{0.05098039,0.05098039,0.05098039}\makebox(0,0)[lt]{\lineheight{1.25}\smash{$digit$}}}%
    \put(0,0){\includegraphics[width=\unitlength,page=5]{rr_tight.pdf}}%
    \put(0.54795737,0.14298401){\color[rgb]{0.03921569,0.03921569,0.03921569}\makebox(0,0)[lt]{\lineheight{1.25}\smash{\,\texttt{\_}}}}%
    \put(0,0){\includegraphics[width=\unitlength,page=6]{rr_tight.pdf}}%
    \put(0.76820604,0.08614565){\color[rgb]{0.05098039,0.05098039,0.05098039}\makebox(0,0)[lt]{\lineheight{1.25}\smash{$space$}}}%
    \put(0,0){\includegraphics[width=\unitlength,page=7]{rr_tight.pdf}}%
    \put(0.10390764,0.22113677){\color[rgb]{0.03921569,0.03921569,0.03921569}\makebox(0,0)[lt]{\lineheight{1.25}\smash{\texttt{,}}}}%
    \put(0,0){\includegraphics[width=\unitlength,page=8]{rr_tight.pdf}}%
  \end{picture}%
\endgroup%